\documentclass[conference]{IEEEtran}
\usepackage{graphicx}
\usepackage{bm}
\usepackage{amsmath}
\usepackage{amssymb}
\usepackage{breqn}
\usepackage{cite}

\ifCLASSOPTIONcompsoc
\usepackage[caption=false,font=normalsize,labelfon
t=sf,textfont=sf]{subfig}
\else
\usepackage[caption=false,font=footnotesize]{subfi
g}
\fi

\newcounter{MYtempeqncnt}
\newtheorem{theorem}{Theorem}
\newtheorem{lemma}{Lemma}

\allowdisplaybreaks[4]

\begin{document}
%
\title{Wireless Communication with
Extremely Large-Scale Intelligent Reflecting Surface}
%
%
%

\author{\IEEEauthorblockN{$\text{Chao~Feng}^*$, $\text{Haiquan~Lu}^*$, $\text{Yong~Zeng}^*\dagger$, $\text{Shi~Jin}^*$, and $\text{Rui~Zhang}^\ddagger$}
\IEEEauthorblockA{*National Mobile Communications Research Laboratory, Southeast University, Nanjing 210096, China\\
$\dagger$Purple Mountain Laboratories, Nanjing 211111, China\\
$\ddagger$Department of Electrical and Computer Engineering, National University of Singapore, Singapore 117583\\
Email: needychao@gmail.com, 
\{haiquanlu, yong\_zeng, jinshi\}@seu.edu.cn} 
elezhang@nus.edu.sg}
\maketitle

\begin{abstract}
  Intelligent reflecting surface (IRS)
  is a promising technology for wireless communications,
  thanks to its potential capability
  to engineer the radio environment.
  However, in practice,
  such an envisaged benefit is attainable 
  only when the passive IRS is of a sufficiently large size,
  for which the conventional uniform plane wave (UPW)-based channel model may become inaccurate.
  In this paper, we pursue
  a new channel modelling and performance analysis
  for wireless communications
  with extremely large-scale IRS (XL-IRS).
  By taking into account
  the variations in signal's amplitude
  and projected aperture across different reflecting elements,
  we derive both lower-
  and upper-bounds of the received signal-to-noise ratio (SNR)
  for the general uniform planar array (UPA)-based XL-IRS.
  Our results reveal that,
  instead of scaling quadratically
  with the increased number of reflecting elements $M$
  as in the conventional UPW model,
  the SNR under the more practically applicable non-UPW model
  increases with $M$ only with a diminishing return and gets saturated eventually.
  To gain more insights,
  we further study the special case of
  uniform linear array (ULA)-based XL-IRS,
  for which a closed-form SNR expression
  in terms of the IRS size
  and transmitter/receiver location is derived.
  This result shows that the SNR
  mainly depends on
  the two geometric angles 
  formed by the transmitter/receiver locations with the IRS, 
  as well as the boundary points of the IRS.
  Numerical results validate our analysis
  and demonstrate the importance of proper channel modelling
  for wireless communications aided by XL-IRS.
\end{abstract}


%
\IEEEpeerreviewmaketitle

\vspace{-0.1cm}
\section{Introduction}
\vspace{-0.1cm}
%
%
%
%


Intelligent reflecting surface (IRS)
is an emerging technology
to achieve cost-effective and energy-efficient wireless communications
by proactively reforming the radio propagation environment
\cite{wu2021intelligentTutorial,wu2019intelligent,huang2019reconfigurable,wu2019towards,Tang2021WirelessCW,di2020smart,lu2021aerial}.
In a nutshell, IRS is a reconfigurable metasurface
consisting of densely arranged low-cost passive elements
and a smart controller.
By adjusting the phase shift
and/or amplitude of the incident signals
on each reflecting element,
the reflected signals
can be added constructively or destructively
at the desired or non-intended receivers,
so as to achieve coverage enhancement,
interference suppression,
security enhancement, enhanced radio localization, etc 
\cite{di2020smart,lu2021aerial}.
Moreover,
IRS avoids costly radio frequency (RF) chains and
operates in a full-duplex mode,
which is thus free of self-interference
and noise amplification.
Besides IRS,
several similar terminologies are also used
in the literature, e.g.,
reconfigurable intelligent surface (RIS)
\cite{huang2019reconfigurable,Tang2021WirelessCW}
and software controllable metasurface (SCS)
\cite{di2020smart,ozdogan2019intelligent}.

Despite of its great potentials,
the promising performance gain brought by IRS
is practically attainable 
only when the size of IRS is sufficiently large \cite{bjornson2019demystifying},
so as to compensate for the double signal attenuation from the transmitter to IRS 
as well as from IRS to the receiver. 
Fortunately,
the appealing features of IRS such as passive reflection without RF chains,
lightweight and conformal geometry
make it possible to 
deploy extremely large-scale IRSs (XL-IRSs) 
in the environment such as the facades of buildings, 
indoor walls
and ceilings.
However, the increased aperture of XL-IRS renders 
that the intended transmitter and/or receiver may not 
be located in the far-field region of the IRS, 
albeit that this generally holds 
for each of its reflecting elements 
due to their much smaller size 
(in the order of carrier wavelength)  
\cite{lu2020how,lu2021communicating}. 
As a result, 
the conventional uniform plane wave (UPW) model 
may become inaccurate for IRS channel modelling.
In this case,
the element-based approach should be adopted 
for more accurate IRS channel modelling and performance analysis, 
by considering the more practical spherical wavefront,
the variations in signal's amplitude 
and angles of arrival/departure (AoA/AoD)
across different reflecting elements.

There have been some preliminary results 
on the mathematical modelling and performance analysis
for wireless communications
without assuming the conventional UPW model,
most of which considered active arrays
\cite{lu2021communicating,hu2018beyond,dardari2020communicating}.
 For example, in \cite{lu2021communicating},
by taking into account the variations in signal's amplitude,
phase and projected aperture over different array elements,
a closed-form expression for the received signal-to-noise ratio (SNR)
was derived for
extremely large-scale array/surface communication,
from which some useful insights were obtained.
In \cite{bjornson2020power},
the power scaling laws and near-field behaviours of IRS
were analyzed for the special case
of two-dimensional (2D) channel modelling
that only considered the azimuth AoA/AoD,
instead of the more general
three-dimensional (3D) modelling
with both azimuth and elevation AoA/AoD.

In this paper,
we study the 3D channel modelling
and performance analysis
for wireless communication
with XL-IRS. By taking into account the variations in signal's
amplitude and projected aperture
across reflecting elements,
tight lower- and upper-bounds of the user received SNR
are derived for the general uniform planar array (UPA)-based XL-IRS.
Our results reveal that instead of scaling quadratically
with the number of reflecting elements (denoted by $M$)
as in the conventional UPW model
\cite{wu2019intelligent,lu2021aerial},
the SNR under the more practical non-UPW model
increases with $M$ only
with a diminishing return and eventually gets saturated. To gain more insights,
we further study the special case
of uniform linear array (ULA)-based XL-IRS,
for which a closed-form SNR expression
in terms of the IRS size and transmitter/receiver location
is derived.
This result shows that the SNR
mainly depends on the two geometric angles 
formed by the transmitter/receiver locations with the IRS, 
as well as the boundary points of the IRS.
Numerical results are provided
to validate our analysis
and demonstrate the necessity of proper channel modelling
for wireless communications aided by XL-IRS.


\vspace{-0.1cm}
\section{System Model}  
\vspace{-0.1cm}

\begin{figure}[ht]
  \vspace{-0.3cm}
  \setlength{\abovecaptionskip}{-0.2cm}
  \setlength{\belowcaptionskip}{-0.2cm}
  \centering
  \includegraphics[width=2.8in]{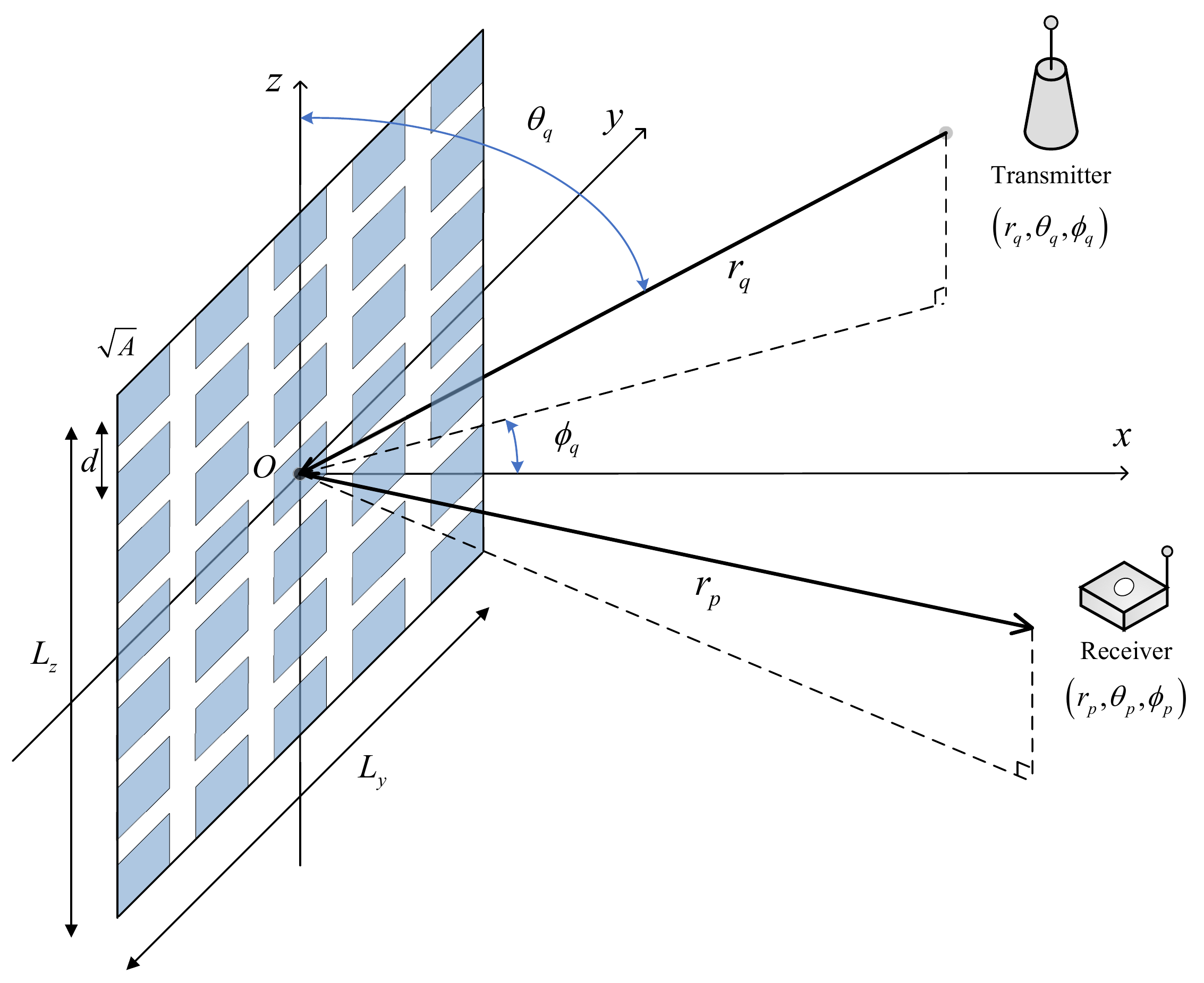}
  \caption{Wireless communication with XL-IRS.}
  \vspace{-0.2cm}
\end{figure}
\vspace{-0.1cm}

As shown in Fig. 1, 
we consider an IRS-aided communication system,
where an XL-IRS is deployed to assist the communication 
between the transmitter and the receiver. 
Without loss of generality, 
the XL-IRS is assumed to be implemented by 
the sub-wavelength discrete UPA.
The number of reflecting elements 
is denoted as $M\gg 1$, 
and the separation between adjacent elements is 
$d\leq \frac{\lambda}2$, 
with $\lambda$ denoting the signal wavelength. 
The physical size of each reflecting element 
is denoted as
$\sqrt{A} \times \sqrt{A}$,
where $\sqrt{A} \leq d$.
We assume that
IRS is placed on the $y$-$z$ plane 
and centered at the origin,
and $M = M_{y}M_{z}$,
where $M_y$ and $M_z$ 
denote the number of reflecting elements 
along the $y$- and $z$-axis, respectively.
Therefore, the total physical size of the IRS
is $L_y \times L_z$,
where $L_y \simeq M_y d$ and $L_z \simeq M_z d$.
 
For notational convenience, 
$M_y$ and $M_z$ are assumed to be odd numbers.
The central location of the $(m_y,m_z)$th reflecting element 
is denoted as
$\mathbf{w}_{m_{y},m_{z}} = [0,m_{y}d,m_{z}d]^T$,
where $m_y = 0,\pm 1,\cdots,\pm (M_y-1)/2$, 
$m_z = 0,\pm 1,\cdots,\pm (M_z-1)/2$.
The transmitter location is denoted by 
$\mathbf{q} = [r_q\Psi_q,r_q\Phi_q,r_q\Omega_q]^T$,
with $\Psi_q \triangleq \sin\theta_q\cos\phi_q$, 
$\Phi_q \triangleq \sin\theta_q\sin\phi_q$, 
and $\Omega_q \triangleq \cos\theta_q$,
where $r_q$ is the distance 
between the transmitter and the center of the XL-IRS, 
and $\theta_q \in [0,\pi]$ and $\phi_q \in [-\frac{\pi}2,\frac{\pi}2]$ 
denote the zenith and azimuth angles, respectively.
The distance between the transmitter 
and the $(m_y,m_z)$th reflecting element is
\begin{align}
  &r_{q,m_{y},m_{z}} = \| \mathbf{w}_{m_{y},m_{z}}-\mathbf{q} \| \notag\\
  & \quad =r_q\sqrt{1-2m_y\varepsilon_q\Phi_q-2m_z\varepsilon_q\Omega_q+(m_y^2+m_z^2)\varepsilon_q^2},
  \label{1}
\end{align}
where $\varepsilon_q \triangleq \frac{d}{r_q}$.
Since the element separation $d$ is on sub-wavelength scale,
we have $\varepsilon_q \ll 1$.

Similarly, denote the receiver location as
$\mathbf{p} = [r_p\Psi_p,r_p\Phi_p,r_p\Omega_p]^T$,
with $\Psi_p \triangleq \sin\theta_p\cos\phi_p$,
$\Phi_p \triangleq \sin\theta_p\sin\phi_p$, 
and $\Omega_p \triangleq \cos\theta_p$,
where $r_p$ is the distance 
between the receiver and the center of the XL-IRS, 
and $\theta_p \in [0,\pi]$ and $\phi_p \in [-\frac{\pi}2,\frac{\pi}2]$ 
are the zenith and azimuth angles, respectively.
The distance between the receiver and the $(m_y,m_z)$th reflecting element is 
\begin{equation}
  r_{p,m_{y},m_{z}}
  =r_p\sqrt{1-2m_y\varepsilon_p\Phi_p-2m_z\varepsilon_p\Omega_p+(m_y^2+m_z^2)\varepsilon_p^2},
  \label{2}
\end{equation}
where $\varepsilon_p \triangleq \frac{d}{r_p} \ll 1$.

For ease of exposition,
we assume that 
the transmitter and receiver each has one antenna 
and their direct link 
is negligible due to severe blockage.
The links 
between the XL-IRS and the transmitter/receiver 
are dominated by the line-of-sight (LoS) path,
by properly placing the IRS in practice.
We focus on IRS implemented by 
using aperture reflecting element 
such as patch element, 
which is of low-profile and especially suitable 
to be mounted on a surface.
For convenience, 
we assume that the aperture efficiency 
is unity so that the \emph{effective aperture} 
of each element is equal to its \emph{physical area}.
By taking into account the variations in signal's amplitude 
and projected aperture across different reflecting elements, 
the channel power gain between the transmitter 
and the $(m_y,m_z)$th element of the XL-IRS 
can be modelled as \cite{lu2021communicating}
\vspace{-0.1cm}
\begin{align}
  &a_{m_{y},m_{z}}(r_q,\theta_q,\phi_q)
  =\underbrace{\frac{1}{4\pi r_{q,m_{y},m_{z}}^2}}_{\text{Free-space pathloss}}
  \underbrace{A\frac{(\mathbf{q}-\mathbf{w}_{m_{y},m_{z}})^T \hat{\mathbf{u}}_x}{\|\mathbf{q}-\mathbf{w}_{m_{y},m_{z}}\|}}_{\text{Projected aperture}} \notag\\
  & \quad =\frac{A\Psi_q}{4\pi r_q^2 [1-2m_y\varepsilon_q\Phi_q-2m_z\varepsilon_q\Omega_q+(m_y^2+m_z^2)\varepsilon_q^2]^{3/2}},
  \label{3}
\end{align}
where $\hat{\mathbf{u}}_x$ is a unit vector 
along the $x$-axis,
which is the normal vector of each IRS element.
Similarly, the channel power gain between the receiver 
and the $(m_y,m_z)$th element of the XL-IRS is
\vspace{-0.1cm}
\begin{align}
  &b_{m_{y},m_{z}}(r_p,\theta_p,\phi_p) \notag\\
  & \quad =\frac{A\Psi_p}{4\pi r_p^2 [1-2m_y\varepsilon_p\Phi_p-2m_z\varepsilon_p\Omega_p+(m_y^2+m_z^2)\varepsilon_p^2]^{3/2}}.
  \label{4}
\end{align}

Denote the channel vector between the transmitter and the XL-IRS 
by $\mathbf{h} \in \mathbb{C}^{M \times 1}$,
whose elements are given by 
\begin{equation}
  h_{m_{y},m_{z}}
  =\sqrt{a_{m_{y},m_{z}}(r_q,\theta_q,\phi_q)} 
  e^{-j\frac{2\pi}{\lambda}r_{q,m_{y},m_{z}}},\forall m_y,m_z.
  \label{5}
\end{equation}

Similarly, denote the channel vector 
between the XL-IRS and the receiver
by $\mathbf{g} \in \mathbb{C}^{M \times 1}$,
with the elements given by
\begin{equation}
  g_{m_{y},m_{z}}
  =\sqrt{b_{m_{y},m_{z}}(r_p,\theta_p,\phi_p)} 
  e^{-j\frac{2\pi}{\lambda}r_{p,m_{y},m_{z}}},\forall m_y,m_z.
  \label{6}
\end{equation}

Further denote by $\theta_{m_y,m_z}$
the phase shift 
introduced by the $(m_y,m_z)$th reflecting element
of the XL-IRS,
and $\mathbf{\Theta} \in \mathbb{C}^{M \times M}$ 
is a diagonal phase shift matrix 
with the diagonal element given by 
$e^{j\theta_{m_y,m_z}}$. 
Then the received signal
can be expressed as
$y=\mathbf{g}^{T}\mathbf{\Theta}{\mathbf{h}} \sqrt{P}s+n$,
where $P$ and $s$ are the transmit power 
and information-bearing symbol, respectively,
and $n\sim\mathcal{CN}(0,\sigma^2)$ is 
the additive white Gaussian noise (AWGN)
at the receiver.

With the optimal phase shifting by the XL-IRS,
i.e., 
$\theta_{m_y,m_z}=\frac{2\pi}{\lambda} r_{q,m_y,m_z}+\frac{2\pi}{\lambda} r_{p,m_y,m_z}$,
the maximum SNR at the receiver can be obtained as
\begin{equation}
  \gamma=\bigg(
    \sum_{m_y=-\frac{M_y-1}2}^{\frac{M_y-1}2} 
    \sum_{m_z=-\frac{M_z-1}2}^{\frac{M_z-1}2}
     |h_{m_y,m_z}||g_{m_y,m_z}| \bigg)^2 
    \bar{P},
    \label{7}
\end{equation}
where $\bar{P} \triangleq \frac{P}{\sigma^2}$.

\vspace{-0.1cm}
\section{Performance Analysis}
\vspace{-0.1cm}

\begin{figure*}[!t]
  \normalsize
  \setcounter{MYtempeqncnt}{\value{equation}}
  \setcounter{equation}{7}

  \begin{equation}
  \begin{aligned}
  \label{8}
  &\gamma=\frac{A^2 \bar{P} \Psi_q \Psi_p}{16\pi^2 r_q^2 r_p^2}\\
  &\times \bigg|\sum_{m_z=-\frac{M_z-1}2}^{\frac{M_z-1}2} \sum_{m_y=-\frac{M_y-1}2}^{\frac{M_y-1}2}
  \frac1{[1-2m_y\varepsilon_q\Phi_q-2m_z\varepsilon_q\Omega_q+(m_y^2+m_z^2)\varepsilon_q^2]^{3/4} 
  [1-2m_y\varepsilon_p\Phi_p-2m_z\varepsilon_p\Omega_p+(m_y^2+m_z^2)\varepsilon_p^2]^{3/4}} \bigg| ^2
  \end{aligned}
  \end{equation}

  \setcounter{equation}{\value{MYtempeqncnt}+1}
  \hrulefill
  \vspace*{4pt}
  \end{figure*}

\begin{figure*}[!t]
  \normalsize
  \setcounter{MYtempeqncnt}{\value{equation}}
  \setcounter{equation}{8}

  \begin{equation}
  \begin{aligned}
  \label{9}
    \gamma
    \simeq 
    \frac{A^2 \bar{P} \Psi_q \Psi_p}{16\pi^2 d^4 r_q^2 r_p^2}  
    \bigg|
    \int_{-\frac{L_z}2}^{\frac{L_z}2}
    \int_{-\frac{L_y}2}^{\frac{L_y}2}
    \frac{\mathrm{d} y \mathrm{d} z}{
    [1-\frac2{r_q} y\Phi_q
    -\frac2{r_q} z\Omega_q+\frac1{r_q^2}(y^2+z^2)]^{3/4}
    [1-\frac2{r_p} y\Phi_p
    -\frac2{r_p} z\Omega_p
    +\frac1{r_p^2}(y^2+z^2)]^{3/4} }
    \bigg|^2 
  \end{aligned}
  \end{equation}

  \setcounter{equation}{\value{MYtempeqncnt}+1}
  \hrulefill
  \vspace*{4pt}
  \end{figure*}
In this section,
performance analysis is carried out 
based on the SNR expression in \eqref{7}.
By substituting \eqref{1}-\eqref{6} into \eqref{7}, 
the resulting SNR 
can be written as \eqref{8}, 
shown at the top of the next page.
Furthermore, by following similar techniques 
as \cite{lu2021communicating} 
and using the fact that 
$\varepsilon_q \ll 1$ and $\varepsilon_p \ll 1$,
the double summation in \eqref{8} can be approximated 
by the corresponding double integral.
As a result,  
the SNR can be expressed in an integral form 
given in \eqref{9}, 
shown at the top of the next page.

\vspace{-0.1cm}
\subsection{SNR Lower- and Upper-Bounds}
\vspace{-0.1cm}

\begin{theorem}
  For the communication aided by an XL-IRS,
  the SNR in \eqref{9} is lower-/upper-bounded by
  \begin{equation}
    f(R_1) \leq \gamma \leq f(R_2),\label{10}
  \end{equation}
  where the function $f(R)$ is defined as \eqref{11}
  shown at the top of the next page,
  and
  $R_1=\frac12 \mathrm{min} \{ L_y,L_z\}$, 
  $R_2=\frac12 \sqrt{L_y^2+L_z^2}$.
\end{theorem}

\begin{figure*}[!t]
  \normalsize
  \setcounter{MYtempeqncnt}{\value{equation}}
  \setcounter{equation}{10}

  \begin{equation}
  \begin{aligned}
  \label{11}
    f(R)
    \triangleq \frac{A^2 \bar{P} \Psi_q \Psi_p}{16\pi^2 d^4 r_q^2 r_p^2}  
    \bigg|
    \int_{0}^{2\pi} \mathrm{d} \zeta
    \int_{0}^{R} 
    \frac{r \mathrm{d} r}{
    (1-\frac{2r}{r_q} \Phi_q\cos\zeta
    -\frac{2r}{r_q} \Omega_q\sin\zeta
    +\frac{r^2}{r_q^2})^{3/4}
    (1-\frac{2r}{r_p} \Phi_p\cos\zeta 
    -\frac{2r}{r_p} \Omega_p\sin\zeta
    +\frac{r^2}{r_p^2})^{3/4} }
    \bigg|^2 
  \end{aligned}
  \end{equation}

  \setcounter{equation}{\value{MYtempeqncnt}+1}
  \hrulefill
  \vspace*{4pt}
  \end{figure*}

\begin{IEEEproof}
  Theorem 1 can be shown 
  by noting that the SNR in \eqref{9} 
  is given by an integral 
  over the rectangular region 
  $L_y \times L_z$ occupied by the XL-IRS.
  By replacing this rectangular region with 
  its inscribed disk and circumscribed disk
  that have radii
  $R_1$ and $R_2$, respectively,
  lower- and upper-bounds in \eqref{10} can be obtained
  as an integral in polar coordinate 
  after a change of variables. 
\end{IEEEproof}

For convenience, we define the distance ratio as
$\rho \triangleq r_q/r_p$.
Without loss of generality,
we may assume that
$0<\rho \leq 1$ due to symmetry.

\begin{lemma}
  If the transmitter and receiver 
  are both located along the boresight of the XL-IRS, 
  i.e., 
  near the $x$-axis
  with $\Phi_q,\Phi_p \ll \frac{r_q}{L_y}$ and 
  $\Omega_q,\Omega_p \ll \frac{r_q}{L_z}$,
  we have
  \begin{equation}\label{12}
    \begin{cases}
      \frac{\rho }{1-\rho^2} \xi^2 \bar{P} G(R_1) \leq \gamma 
      \leq \frac{\rho }{1-\rho^2} \xi^2 \bar{P} G(R_2),& 0<\rho<1\\
      \\
      \gamma=\frac{\xi^2 \bar{P}}{\pi^2} \arctan^2 
      \frac{(\frac{L_y}{2 r_q})(\frac{L_z}{2 r_q})}
      {\sqrt{(\frac{L_y}{2 r_q})^2+(\frac{L_z}{2 r_q})^2+1}},& \rho=1
    \end{cases}
\end{equation}
where $\xi \triangleq \frac{A}{d^2}$ with $\rho<1$
is the \emph{array occupation ratio} \cite{lu2021communicating}, 
and the function $G(R)$ is defined as
\begin{align}
  G(R) 
  &\triangleq 
  \bigg[F(\frac12 \arctan \frac{\sqrt{1-\rho^2}}{\rho}|2)\notag\\
  &-F(\frac12 \arctan(\frac{\sqrt{1-\rho^2}}{\rho}
  \cos(\arctan \frac{R}{r_q}))|2)\bigg]^2,
  \label{13}
\end{align}
and $F(\vartheta|k)=\int_0^{\vartheta} \frac{1}{\sqrt{1-k\sin^2\beta}}\mathrm{d} \beta$
is the incomplete Elliptic Integral of the First Kind \cite{luke1968approximations}. 
\end{lemma}
\begin{IEEEproof}
  Please refer to Appendix A.
\end{IEEEproof}

\begin{lemma}
  Under the same condition as Lemma 1,
  the asymptotic SNR aided by the XL-IRS is
  \begin{align} \label{14}
    \hspace{-0.2cm}
  \lim_{L_y,L_z \rightarrow \infty} \gamma =
  \begin{cases}
    \frac{\rho}{1-\rho^2} \xi^2 \bar{P}
    \bigg[F(\frac12 \arctan \frac{\sqrt{1-\rho^2}}{\rho}|2)\bigg]^2,&0<\rho<1\\
    \\
    \frac{\xi^2 \bar{P}}{\pi^2} \times (\frac{\pi}2)^2
    =\frac{\xi^2 \bar{P}}{4},&\rho=1
  \end{cases}
  \end{align} 
\end{lemma}
\begin{IEEEproof}
  For $0<\rho<1$,
  as $L_y,L_z \rightarrow \infty$,
  the radii of
  the inscribed disk and the circumscribed disk
  also go to infinity,
  i.e., $R_1,R_2 \rightarrow \infty$.
  It then follows from \eqref{12} and \eqref{13} 
  that both lower- and upper-bounds of the SNR 
  approach to the same value.
  Therefore,
  the first case of \eqref{14} follows
  according to the Squeeze Theorem \cite{thomas1961calculus}.
  Besides, for $\rho=1$,
  the resulting SNR
  can be easily obtained 
  by letting $L_y,L_z \rightarrow \infty$
  in the second case of \eqref{12}.
\end{IEEEproof}

As a comparison,
the SNR under the commonly used
UPW model for the same system setup is given by
\cite{wu2021intelligentTutorial,lu2021aerial}
\begin{equation} \label{15}
  \gamma_{UPW}
  =\frac{\beta_0^2 \bar{P}}{r_q^2 r_p^2} M^2,
\end{equation}
where $\beta_0$ is the channel gain
at the reference distance of $1\,$m.
The result \eqref{15} is known as 
the square power scaling law 
for IRS-assisted communication, 
which is valid when both $r_q$ and $r_p$ 
are sufficiently large as compared to the IRS dimension 
(i.e., $M$ is moderately large), 
under which the far-field propagation model 
holds for both the whole IRS 
as well as each of its individual elements. 
However, if $M$ goes even larger, 
the above result cannot hold anymore as 
the square power scaling law implies that 
the SNR would increase unboundedly, 
which is obviously impractical. 
In contrast, our new result in Lemma 2 reveals that 
under the practical non-UPW model, 
the SNR will increase with $M$,
but with a diminishing return, 
and eventually approach to a constant 
that only depends on the distance ratio $\rho$, 
and the array occupation ratio $\xi$.  
\begin{figure}[ht]
  \vspace{-0.6cm}
  \setlength{\abovecaptionskip}{-0.2cm}
  \setlength{\belowcaptionskip}{-0.2cm}
  \centering
  \includegraphics[width=2.8in]{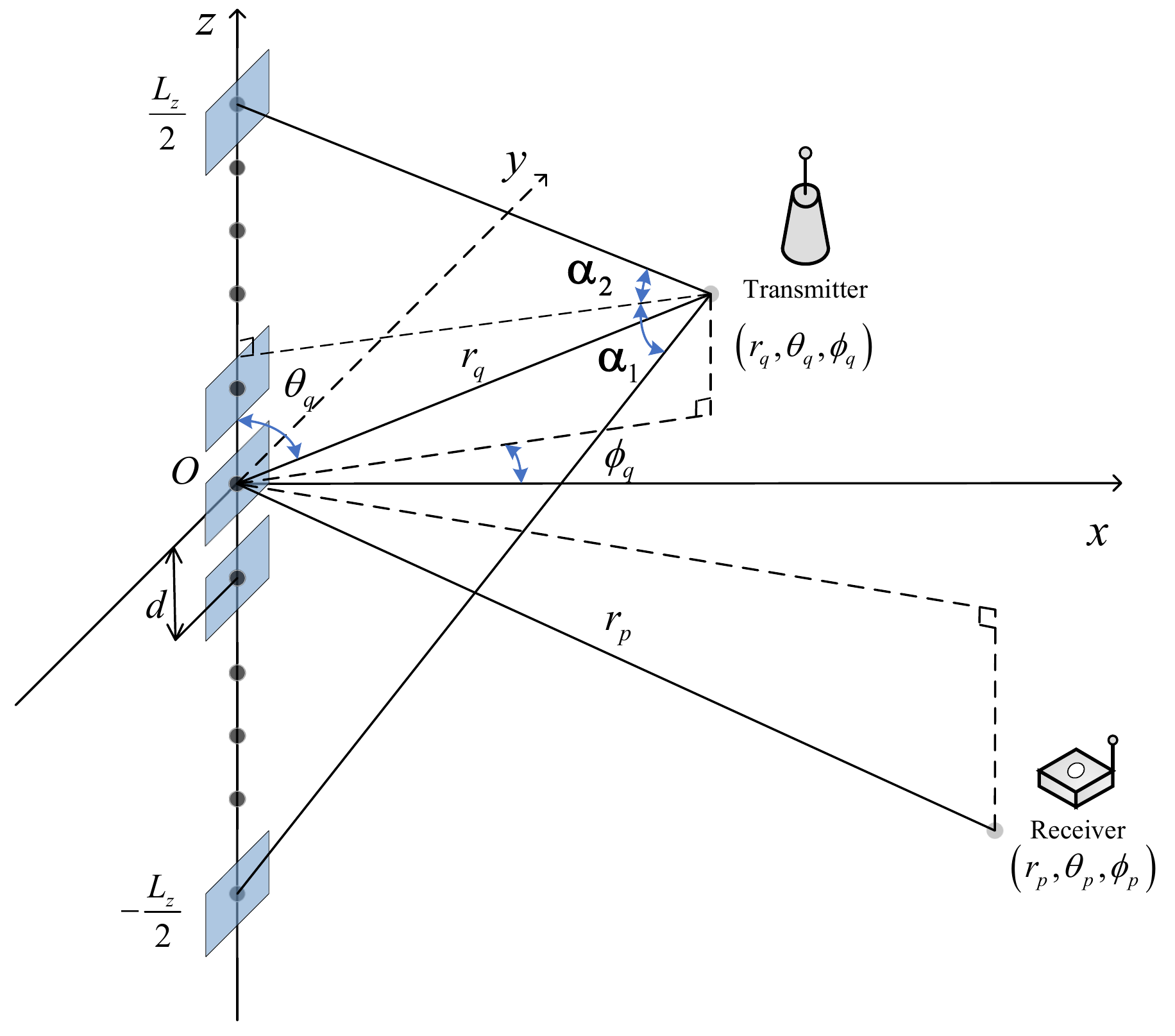}
  \caption{Wireless communication with ULA-based XL-IRS.}
  \vspace{-0.6cm}
\end{figure}

\begin{figure*}[!t]
  \normalsize
  \setcounter{MYtempeqncnt}{\value{equation}}
  \setcounter{equation}{15}

  \begin{equation}
  \begin{aligned}
  \label{16}
  \gamma
  \simeq \frac{A^2 \bar{P} \Psi_q \Psi_p}{16\pi^2 d^2 r_q^2 r_p^2}
  \bigg|
  \int_{-\frac{L_z}2}^{\frac{L_z}2} 
  \frac{\mathrm{d} z}
  {[(1-\frac2{r_q} z\cos\theta_{q}+\frac{z^2}{r_q^2})
  (1-\frac2{r_p} z\cos\theta_p+\frac{z^2}{r_p^2})]^{3/4}} \bigg|^2
  \end{aligned}
  \end{equation}

  \setcounter{equation}{\value{MYtempeqncnt}+1}
  \hrulefill
  \vspace*{4pt}
  \end{figure*}

  \vspace{-0.03cm} 
\subsection{ULA-based XL-IRS}
\vspace{-0.1cm} 

To gain more insights, 
we consider the special case of ULA-based XL-IRS, 
where $M_y=1$ and $M_z=M$.
In this case,
by letting $y=0$ and $\mathrm{d} y=d$,
the SNR expression in \eqref{9}
reduces to \eqref{16}
shown at the top of the next page.

\begin{lemma}
  For the communication aided by a ULA-based XL-IRS, 
  when $r_q \ll r_p$ (i.e., $\rho \rightarrow 0$), 
  the SNR in \eqref{16} can be expressed as 
  \begin{equation} \label{17}
    \gamma=\frac{A^2 \bar{P} \Psi_p \cos\phi_q }{4\pi^2d^2r_{p}^2} 
    \bigg[F(\frac{\alpha_1}{2}| 2)+F(\frac{\alpha_2}{2}| 2)\bigg]^2,
  \end{equation}
  where $\alpha_1=\arctan\frac{L_z/2+r_q\cos\theta_q}{r_q\sin\theta_q}$
  and $\alpha_2=\arctan\frac{L_z/2-r_q \cos\theta_q}{r_q\sin\theta_q}$.
\end{lemma}

\begin{IEEEproof}
  Please refer to Appendix B.
\end{IEEEproof}

Note that the condition $r_q\ll r_p$ 
in Lemma 3 corresponds to the typical IRS deployment scenario, 
where it has been shown that 
IRS should be deployed closer to either the transmitter 
or receiver for SNR maximization 
\cite{wu2021intelligentTutorial,lu2021aerial}.
Lemma 3 shows that with a ULA-based XL-IRS, 
the IRS size $L_z$ affects the SNR 
via the two geometric angles, $\alpha_1$ and $\alpha_2$,
which are the angles 
formed by the line segments
connecting the transmitter location
and its projection to the IRS,
as well as the two ends of the IRS,
as shown in Fig. 2.
In particular, $\alpha_1+\alpha_2$ 
is termed as the \emph{angular span}
\cite{lu2020how}.
It is not difficult to see that 
both $\alpha_1$ and $\alpha_2$ increase with the IRS size $L_z$
and decrease with the distance $r_q$.
Since the Elliptic Integral function 
$F(\vartheta|2)$ 
monotonically increases with $\vartheta$,
the SNR $\gamma$ in \eqref{17} increases with $L_z$
but decreases with $r_q$, 
as expected.
Furthermore, different from 
the conventional square power scaling law obtained 
based on the UPW model 
where the SNR increases unboundedly with the IRS size
\cite{wu2019intelligent,lu2021aerial}, 
Lemma 3 shows that under the non-UPW model, 
the SNR increases with $L_z$ with a diminishing return. 
In particular, as $L_z \rightarrow \infty$, 
we have $\alpha_1=\alpha_2=\frac{\pi}2$, 
which leads to the following result.

\begin{lemma}
  Under the same condition as Lemma 3, 
  the asymptotic SNR in  the case of ULA-based XL-IRS is 
  \begin{align} \label{18}
    \lim_{L_z \rightarrow \infty} \gamma
    & =
    \frac{A^2 \bar{P} \Psi_p \cos\phi_q }{\pi^2 d^2 r_p^2} 
    \bigg[F(\frac{\pi}{4}| 2) \bigg]^2 \notag\\
    &= 1.7188 \times 
    \frac{A^2 \bar{P} \Psi_p}{\pi^2 d^2r_p^2} \cos\phi_q.
  \end{align}
\end{lemma}

\vspace{-0.1cm}
\section{Numerical Results}
\vspace{-0.1cm}
In this section,
numerical results are provided to
validate our theoretical analysis,
and also compare our proposed model 
with the conventional UPW model.
Unless otherwise stated,
the signal wavelength is $\lambda=0.125\,$m,
the element seperation is set as $d=\frac{\lambda}5$,
and the element size is $A=(\frac{d}2)^2$,
which corresponds to the array occupation ratio 
$\xi=\frac14$.

\begin{figure}[ht]
  \vspace{-0.6cm}
  \setlength{\abovecaptionskip}{-0.2cm}
  \setlength{\belowcaptionskip}{-0.2cm}
  \centering
  \includegraphics[width=3.0in]{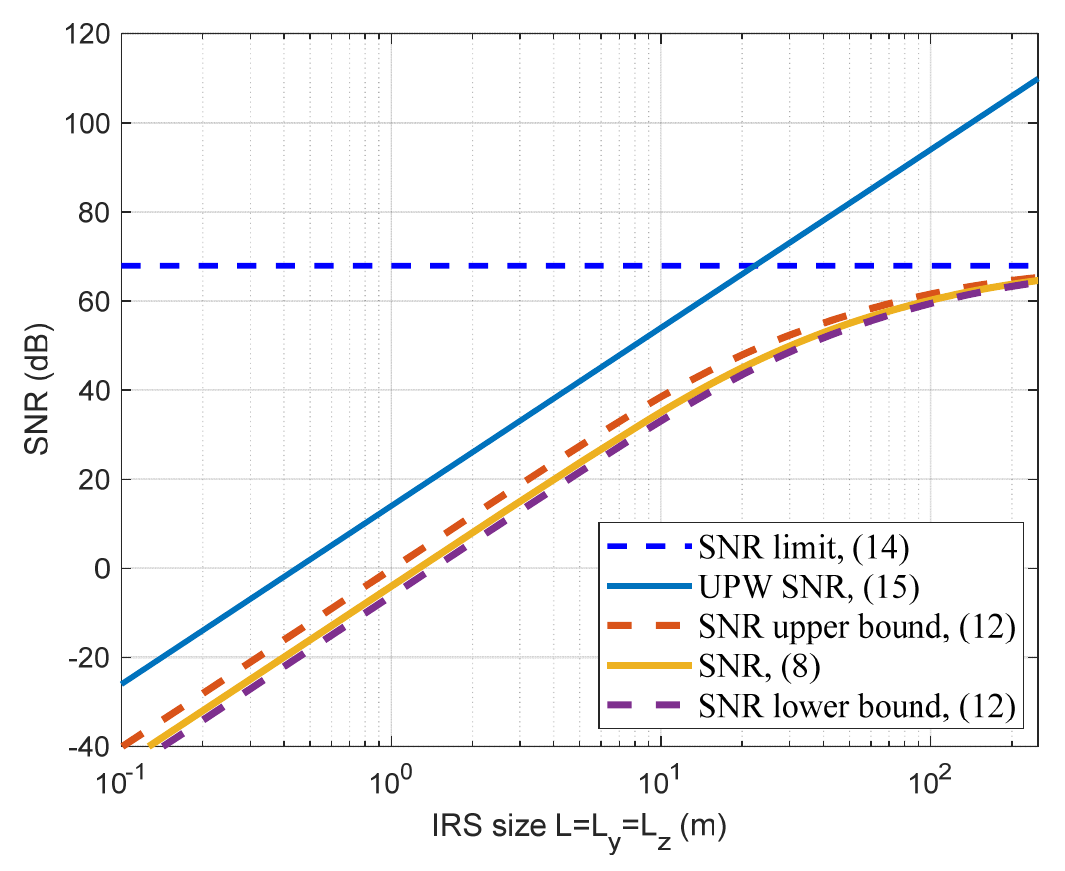}
  \caption{SNR versus IRS size for UPA-based XL-IRS.}
  \vspace{-0.6cm}
\end{figure}

Fig. 3 plots the SNR 
versus IRS size for square UPA-based XL-IRS, i.e., $L=L_y=L_z$.
The results based on the summation in \eqref{8},
closed-form lower- and upper-bounds in \eqref{12},
the asymptotic value in \eqref{14},
and that under 
the conventional UPW model in \eqref{15} are compared.
The transmit SNR is $\bar{P}=90\,$dB, and
the locations of the transmitter and receiver are 
$\mathbf{q}=[10,0,0]^T\,$m 
and $\mathbf{p}=[100,0,0]^T\,$m, respectively.
It is firstly observed that 
the derived closed-form bounds in Lemma 1
are quite accurate for the SNR prediction 
in XL-IRS aided communications.
Furthermore, as the IRS size $L$ increases,
the SNR approaches to a constant,
which verifies  
the theoretical result in Lemma 2.
Besides,
it is observed that
the conventional UPW model \eqref{15} in general
tends to
over-estimate the SNR value,
and as the IRS size 
goes beyond a certain threshold,
the two models exhibit drastically different scaling laws,
i.e., approaching to a constant value 
versus increasing unboundedly.

\begin{figure}[ht]
  \vspace{-0.4cm}
  \setlength{\abovecaptionskip}{-0.2cm}
  \setlength{\belowcaptionskip}{-0.2cm}
  \centering
  \includegraphics[width=3.0in]{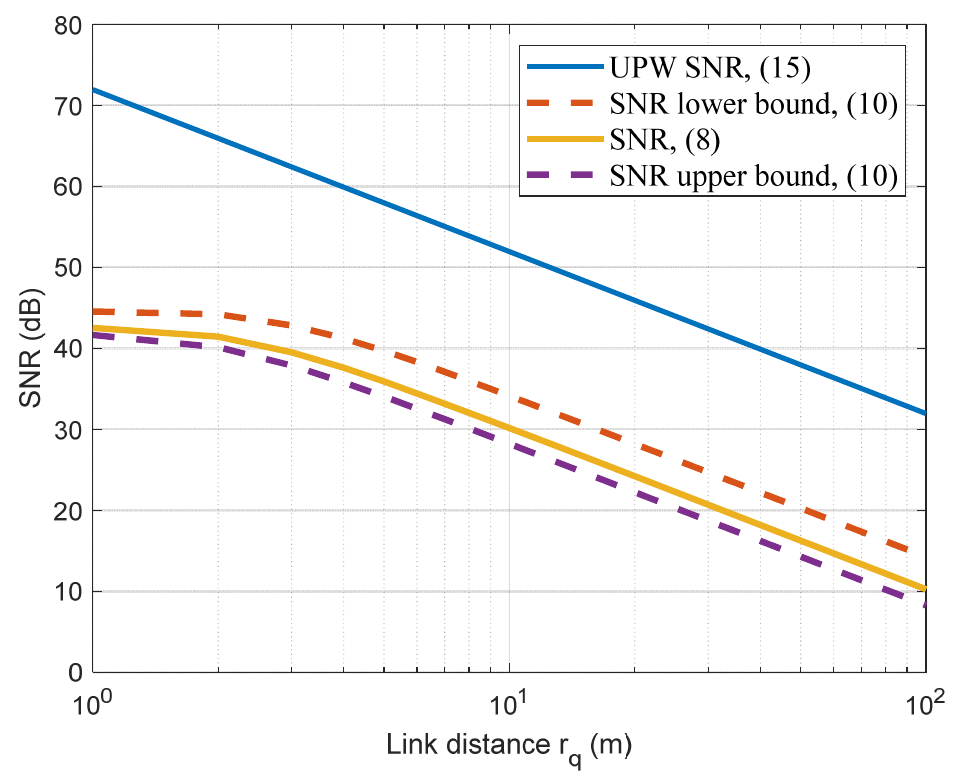}
  \caption{SNR versus link distance $r_q$ for UPA-based XL-IRS.}
  \vspace{-0.2cm}
\end{figure}

Fig. 4 plots the SNR  
versus the link distance 
between the transmitter and IRS $r_q$
for UPA-based XL-IRS,
which has size $L_y=L_z=5\,$m.
The direction of the transmitter is
$(\theta_q,\phi_q)=(\frac{\pi}3,\frac{\pi}6)$
and the location of the receiver is
$(r_p,\theta_p,\phi_p)=(200\,\mathrm{m},\frac{3\pi}4,-\frac{\pi}5)$,
respectively.
The transmit SNR is $\bar{P}=100\,$dB.
It is observed that
the bounds given in Theorem 1 are tight, 
and the conventional UPW model over-estimates the SNR values.   
In particular,
for relatively small
link distance $r_q$,
different SNR scalings versus $r_q$ 
are observed for the conventional UPW model 
and the newly considered non-UPW model, 
which leads to significantly different SNR values, 
e.g., by a difference about $25\,$dB for $r_q=2\,$m.  

\begin{figure}[ht]
  \vspace{-0.6cm}
  \setlength{\abovecaptionskip}{-0.2cm}
  \setlength{\belowcaptionskip}{-0.2cm}
  \centering
  \includegraphics[width=3.0in]{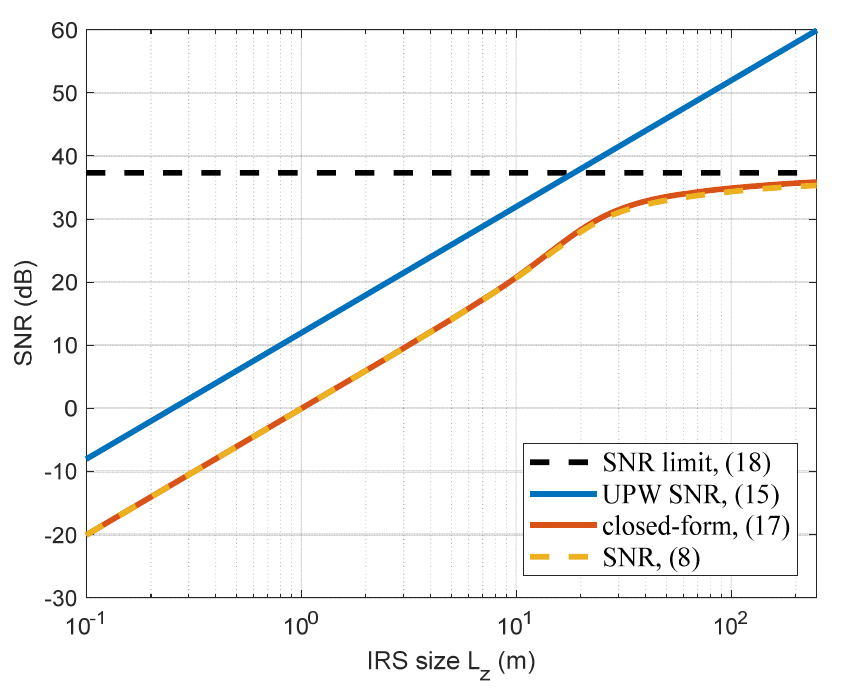}
  \caption{SNR versus IRS size for ULA-based XL-IRS.}
  \vspace{-0.5cm}
\end{figure}

For ULA-based XL-IRS,
Fig. 5 shows the SNR
versus the IRS size $L_z$
based on the summation \eqref{8},
the derived closed-form expression \eqref{17},
the asymptotic limit \eqref{18},
and that under the conventional UPW model \eqref{15}.
The locations of the transmitter and receiver are
$(r_q,\theta_q,\phi_q)=(10\,\mathrm{m},\frac{\pi}3,\frac{\pi}6)$
and
$(r_p,\theta_p,\phi_p)=(100\,\mathrm{m},\frac{3\pi}4,-\frac{\pi}5)$,
respectively,
and the transmit SNR is $\bar{P}=120\,$dB.
It is firstly observed that 
the closed-form expression \eqref{17} and 
the asymptotic limit \eqref{18} 
match well with the actual values. 
Furthermore, similar to Fig. 3, 
there exists significant gap between
the conventional UPW model and 
our considered model.
In particular,
as $L_z$ becomes sufficiently large,
while the SNR under the UPW model increases unboundedly, 
that under our considered model 
approaches to a constant value specified in \eqref{18}.  
This again demonstrates the importance 
of proper channel modelling 
for communications aided by XL-IRS.

\vspace{-0.1cm}
\section{Conclusions}
\vspace{-0.1cm}
This paper studied the mathematical modelling 
and performance analysis of 
wireless communication aided by XL-IRS. 
By taking into account the variations in signal's amplitude 
and projected aperture across different reflecting elements, 
we firstly derived tight lower- and upper-bounds 
of the receiver SNR 
for the general UPA-based XL-IRS. 
To gain more insights, 
the special case of ULA-based XL-IRS was also considered, 
for which a closed-form SNR expression 
in terms of the ULA size and 
transmitter/receiver locations was derived. 
Numerical results verified our theoretical analysis 
and demonstrated the importance of proper channel modelling 
for wireless communications aided by XL-IRS.

\vspace{-0.1cm}
\appendices
\section{Proof of Lemma 1}
\vspace{-0.1cm}

The double integral in \eqref{11} 
reduces to the following form under
the assumption of
$\Phi_q,\Phi_p \ll \frac{r_q}{L_y}$ and 
  $\Omega_q,\Omega_p \ll \frac{r_q}{L_z}$.
\begin{align}\label{19}
  I_1
  &\simeq
  \int_0^{2\pi} \mathrm{d} \zeta 
  \int_0^{R} 
  \frac{r \mathrm{d} r}
  {[(1+\frac{r^2}{r_q^2})(1+\frac{r^2}{r_p^2})]^{3/4}} \notag\\
  &=2\pi r_q^2
  \int_0^{\frac{R}{r_q}} 
  \frac{r \mathrm{d} r}
  {[(r^2+1)(\rho^2 r^2+1)]^{3/4}}.
\end{align}

By letting $r=\tan\alpha$,
\eqref{19} can be simplified as 
\begin{equation}\label{20}
  I_1=2\pi r_q^2
  \int_0^{\arctan \frac{R}{r_q}} 
  \frac{\sin\alpha \mathrm{d}\alpha}
  {[\rho^2+(1-\rho^2)\cos^2\alpha]^{3/4}}.
\end{equation}
We first consider the case of $0<\rho<1$.  
By letting $v=\frac{\sqrt{1-\rho^2}}{\rho}\cos\alpha$,
$I_1$ can be further expressed as 
\begin{align}\label{21}
  I_1 
  &= 2\pi r_q^2 \frac{\rho^{-1/2}}{\sqrt{1-\rho^2}}
  \int_{\frac{\sqrt{1-\rho^2}}{\rho}\cos(\arctan \frac{R}{r_q})}^{\frac{\sqrt{1-\rho^2}}{\rho}} 
  \frac{\mathrm{d} v}
  {(v^2+1)^{3/4}}.\notag\\
  &=4\pi r_q^2 \frac{\rho^{-1/2}}{\sqrt{1-\rho^2}} 
  \int_{\arctan(\frac{\sqrt{1-\rho^2}}{\rho}\cos(\arctan \frac{R}{r_q}))}^{\arctan\frac{\sqrt{1-\rho^2}}{\rho}} 
  \frac{\mathrm{d} \frac{\varphi}2}{\sqrt{1-2\sin^2 \frac{\varphi}2}},
\end{align}
where the last equality follows by a change of variable as
$v=\tan\varphi$.
According to the definition of 
incomplete Elliptic Integral of the First Kind,
\eqref{21} can be written as
\begin{equation}\label{22}
  I_1 
  = 4\pi r_q^2 \frac{\rho^{-1/2}}{\sqrt{1-\rho^2}} \sqrt{G(R)},
\end{equation}
where $G(R)$ is defined in \eqref{13}.

By substituting \eqref{22} into \eqref{11} and with Theorem 1,
the first case of \eqref{12} in Lemma 1 can be obtained.

For the special case of $\rho=1$,
i.e., $r_q=r_p$,
\eqref{12} can be obtained from the integral in \eqref{9},
where under the condition of Lemma 1, we have
\begin{align}\label{23}
  I_2
  &=\int_{-\frac{L_z}2}^{\frac{L_z}2}
  \int_{-\frac{L_y}2}^{\frac{L_y}2}
  \frac{\mathrm{d} y \mathrm{d} z}{
  (1+\frac{y^2}{r_q^2}+\frac{z^2}{r_q^2})^{3/2}}\notag\\
  &=4r_q^2\arctan
  \frac{(\frac{L_y}{2 r_q})(\frac{L_z}{2 r_q})}
  {\sqrt{(\frac{L_y}{2 r_q})^2+(\frac{L_z}{2 r_q})^2+1}}.
\end{align}

By substituting \eqref{23} into \eqref{9},
the second case of \eqref{12} then follows.

\vspace{-0.1cm}
\section{Proof of Lemma 3}
\vspace{-0.1cm}
By applying a change of variable as $t=\frac{z}{r_q}$,
the integral in \eqref{16} can be expressed as
\begin{equation} \label{24}
  I=\int_{-\frac{L_z}{2 r_q}}^{\frac{L_z}{2 r_q}} \frac{r_q\mathrm{d} t}
  {[(1-2t\cos\theta_{q}+t^2)(1-2\rho t\cos\theta_{p}+\rho^2 t^2)]^{3/4}}.
\end{equation}

By letting $u=\frac{t-\cos\theta_q}{\sin\theta_q}$, 
$I$ can be further written as 
\begin{equation}\label{25}
  I=\frac{r_q}{\sqrt{\sin\theta_q}}
  \int_{u_1}^{u_2} 
  \frac{\mathrm{d} u}
  {(u^2+1)^{3/4} 
  [(u^2+1)X
  +Yu+1+Z]^{3/4}},
\end{equation}
where 
\begin{align*}
  &u_1=-\frac{L_z/2+r_q\cos\theta_q}{r_q\sin\theta_q},
  u_2=\frac{L_z/2-r_q\cos\theta_q}{r_q\sin\theta_q},\\
  &X=\rho^2 \sin^2\theta_q,\\
  &Y=2\rho^2\sin\theta_q\cos\theta_q-2\rho\sin\theta_q\cos\theta_p,\\
  &Z=\rho^2\cos^2\theta_q-2\rho\cos\theta_q\cos\theta_p-\rho^2 \sin^2\theta_q.
\end{align*}

Then by letting $u=\tan\varphi$, \eqref{25} can be simplified as 
\begin{equation}\label{26}
  I=\frac{r_q}{\sqrt{\sin\theta_q}}
  \int_{\varphi_1}^{\varphi_2} 
  \frac{\cos\varphi \mathrm{d} \varphi}
  {[X+Y\sin\varphi\cos\varphi+(1+Z)\cos^2\varphi]^{3/4}},
\end{equation}
where $\varphi_1=-\arctan\frac{L_z/2+r_q\cos\theta_q}{r_q\sin\theta_q}$
and $\varphi_2=\arctan\frac{L_z/2-r_q\cos\theta_q}{r_q\sin\theta_q}$.

Under the condition of Lemma 3, we have $\rho \ll 1$, 
and hence $X,Y,Z \ll 1$.
Thus, \eqref{26} reduces to
  \begin{equation}\label{27}
    I\simeq \frac{r_q}{\sqrt{\sin\theta_q}}
    \int_{\varphi_1}^{\varphi_2} \frac{\mathrm{d} \varphi}{\sqrt{\cos\varphi}}.
  \end{equation}

  Based on the definition of 
  incomplete Elliptic Integral of the First Kind, 
  we have
  \begin{equation}\label{28}
   I=\frac{2r_q}{\sqrt{\sin\theta_q}} \bigg[F(-\frac{\varphi_1}{2}| 2)+F(\frac{\varphi_2}{2}| 2)\bigg].
  \end{equation}

  With the identities that 
  $\alpha_1=-\varphi_1$ and $\alpha_2=\varphi_2$,
  and by substituting \eqref{28} into \eqref{16}, 
  the proof of Lemma 3 is completed.

\vspace{-0.2cm}
\bibliographystyle{IEEEtran}
\bibliography{ref}

%









\end{document}